\documentclass[12pt]{article}
\usepackage{epsf}
\usepackage{pazha}
\tightenlines

\def\ergs{erg~s$^{-1}$}
\def\*{$^{*}$}
\def\a{$^{\mbox{\small a}}$}
\def\b{$^{\mbox{\small b}}$}
\def\c{$^{\mbox{\small c}}$}
\def\d{$^{\mbox{\small d}}$}
\def\INTEGRAL{\hbox{INTEGRAL}}

\def\igr{\hbox{IGR\,J18462-0223}}
\def\etal{{et~al.}}
\sloppypar

\begin{document}
\baselineskip 21pt
\noindent
{\sl will be published in Astronomy Letters, 2010, 36, 533.\/}

\vspace{2cm}

\title{\Large\bf NEW FAST X-RAY TRANSIENT IGR\,J18462-0223 DISCOVERED
BY THE {\bf INTEGRAL\/}\\ OBSERVATORY}

\author{\bf S.A. Grebenev\affilmark{1*}, R.A. Sunyaev\affilmark{1,2}} 

\affil{
$^1$ {\it Space Research Institute, RAS,  Profsoyuznaya 84/32, 117997 Moscow, Russia}\protect\\ 
$^2$ {\it MPI f\"ur Astrophysik, Karl-Schwarzschild-Str. 1,
    D-85741 Garching, Germany}
} 

\vspace{2mm}
\received{March 16, 2010}

\sloppypar 
\vspace{2mm}
\noindent
Details of the discovery of a new X-ray source, \igr, on October
12, 2007, during a short (several hours), intense ($\sim 35$
mCrab at the peak) outburst of hard radiation by the
\mbox{IBIS/ISGRI} gamma-ray telescope onboard the \INTEGRAL\
observatory are given. The detection of another earlier outburst
from this source occurred on April 28, 2006, in the archival
data of the telescope is reported. We present the results of the
source's localization and our spectral/timing analysis of the
observational data. The source may turn out to be yet another
representative of the continuously growing population of fast
X-ray transients, which are the focus of attention because of
the identification of their optical counterparts with early-type
supergiants.

\noindent
{\bf Key words: \/} X-ray sources, fast transients, X-ray
pulsars, accretion from stellar wind.

\vfill
\noindent\rule{8cm}{1pt}\\
{$^*$ E-mail: $<$sergei@hea.iki.rssi.ru$>$}

\clearpage

\section*{INTRODUCTION}
\noindent
In October-November 2007, the \INTEGRAL\ international orbital
gamma-ray observatory was measuring the spectral shape and
distribution profile of the Galactic ridge hard X-ray emission
in the Scutum Arm tangent. During the observation on October 12,
the \mbox{IBIS/ISGRI} gamma-ray telescope detected a hitherto
unknown X-ray source named \igr\ (Grebenev et al. 2007). The
emission from the source was detected with confidence for at
least 5 h (out of the 8 h during which it was in the telescope's
field of view). The source was detected neither before (27 h
earlier) nor after (12 h later) this observation.

The communication by Grebenev et al. (2007) contained only brief
information about the discovery of \igr. Here, we present the
results of a more detailed spectral/timing analysis of the
INTEGRAL data obtained during this outburst of the source and
report the detection of another earlier (occurred on April 28,
2006) X-ray outburst from it.

The source is interesting in that it could be a new
representative of the continuously growing population of fast
X-ray transients, which are the focus of attention because of
the identification of their optical counterparts with early-type
supergiants (Smith 2004; Negueruela et al. 2006b). The fast
transients are characterized by short (less than one day)
intense (with a peak luminosity of $\sim 10^{36}-10^{37}$ \ergs)
X-ray outbursts separated by long (tens and hundreds of days)
periods of quiescence, during which their luminosity falls by
three or four orders of magnitude. The population already
numbers at least eleven sources: IGR\,J08408-4503 (G\"otz et
al. 2007), IGR\,J11215-5952 (Lubinski et al. 2005; Negueruela et
al. 2005; Sidoli et al. 2007), IGR\,J16465-4507 (Lutovinov et
al. 2004, 2005; Zurita Heras and Walter 2004; Smith 2004),
IGR\,J16479-4514 (Molkov et al. 2003; Sguera et al. 2006; Chaty
et al. 2008), XTE\,J1739-302 (Smith et al. 1998, 2006;
Negueruela et al. 2006a), AX\,J1749.1-2733 (Sakano et al. 2002;
Grebenev and Sunyaev 2007; Karasev et al. 2009),
IGR\,J17544-2619 (Sunyaev et al. 2003; Grebenev et al. 2004;
in't Zand 2005; Pellizza et al. 2006), SAX\,J1818.6-1703 (in't
Zand et al. 1998; Grebenev and Sunyaev 2005; Sguera et al. 2006;
Negueruela and Smith 2006), AX\,J1841.0-0536 (Bamba et al. 2001;
Sguera et al.  2006; Negueruela et al. 2006b), AX\,J1845.0-0433
(Yamauchi et al. 1995; Sguera et al. 2007),
and IGR\,J18483-0311 (Chernyakova et al. 2003; Molkov et
al. 2004; Rahoui and Chaty 2008). It is beyond doubt that the
compact object in these sources is a neutron star with a strong
magnetic field (pulsar) accreting matter from a supergiant's
dense stellar wind. X-ray pulsations have already been detected
from five sources of the group. The remaining sources may be
considered pulsars, because their radiation spectra are
similar. The only question is why the accretion is episodic and
so short in duration.

However, irrespective of the answer to this question, the
discovery of fast transients itself provided a wealth of
information about the variety of observational appearances of
high-mass X-ray binaries with supergiants. This discovery also
advances noticeably our understanding of the problem of
observing a very limited number of (quasi)persistent sources of
this type compared to theoretical expectations (Illarionov and
Sunyaev 1975; Grebenev and Sunyaev 2007).

\section*{OBSERVATIONS}
\noindent

The \INTEGRAL\ observatory (Winkler et al. 2003) was placed in a
high-apogee orbit by a PROTON launcher on October 17, 2002
(Eismont et al. 2003).  It is equipped with four telescopes
capable of carrying out simultaneous gamma-ray, X-ray, and
optical observations. This work is based on data from the ISGRI
detector (Lebrun et al. 2003) of the IBIS gamma-ray telescope
(Ubertini et al. 2003)\footnote{The JEM-X monitor onboard
INTEGRAL could give data in a softer X-ray band than that of
ISGRI, but IGR J18462-0223 was far from the center of its field
of view (narrower than the IBIS one) and was observed with a low
sensibility.}. This telescope, which uses the principle of a
coded aperture, allows one to image the sky in a 30\deg
$\times$30\deg\ field of view (the fully coded zone is 9\deg
$\times$9\deg) with an angular resolution of 12\arcmin\ (FWHM)
and to investigate the properties of the detected point
sources. ISGRI is a position-sensitive detector that consists of
128$\times$128 CdTe semiconductor elements efficiently operating
in the energy range 18--200 keV. The total area of the elements
reaches 2620 cm$^2$; the effective area for sources at the
center of the field of view is $\sim$ 1100 cm$^2$ (one half of
the detector is shadowed by the opaque aperture elements). The
detector provides an energy resolution $\Delta E/E\sim 7$\%
(FWHM).

The observations of the Scutum Arm tangent in mid-October 2007,
during which IGR J18462-0223 was discovered, were carried out in
the accordance with the proposal of R.A. Sunyaev. The
observatory was successively pointed at points with a fixed
Galactic longitude $l\simeq24$\deg\ ($l\simeq22$\deg\ at the next
passage) and a latitude changing from $b=30$\deg\ to $-30$\deg\
at $2$\deg\ steps.

Another outburst of the source, which was revealed already after
its discovery when the archival data were analyzed, was recorded
in late April 2006. At this time, INTEGRAL was scanning the
Galactic plane within the framework of its Core observational
program (Winkler et al. 2003). The observatory was pointed at
points in the band $b=\pm4$\deg\ at $2$\deg\ steps.  Depending
on the pointing, the exposure efficiency for \igr\ (its Galactic
coordinates are $l\simeq30$\fdg$2$, $b=0$\fdg$06$) changed. The
duration of each pointing was $\sim2000$~s in both cases.

The observational data were analyzed using the software
developed for the IBIS/ISGRI telescope at the Space Research
Institute, the Russian Academy of Sciences. Its general
description can be found in Revnivtsev et al. (2004). In our
spectral analysis, we used the response matrix of the standard
OSA package, which proved to be efficient in fitting the spectra
of the Crab Nebula. The spectrum of the Nebula was assumed to be
$dN(E)/dE=10\, E^{-2.1}$ phot cm$^{-2}$ s$^{-1}$ keV$^{-1}$,
where the energy $E$ is given in keV.
\begin{figure}[t]
\epsfxsize=17cm 
\epsffile{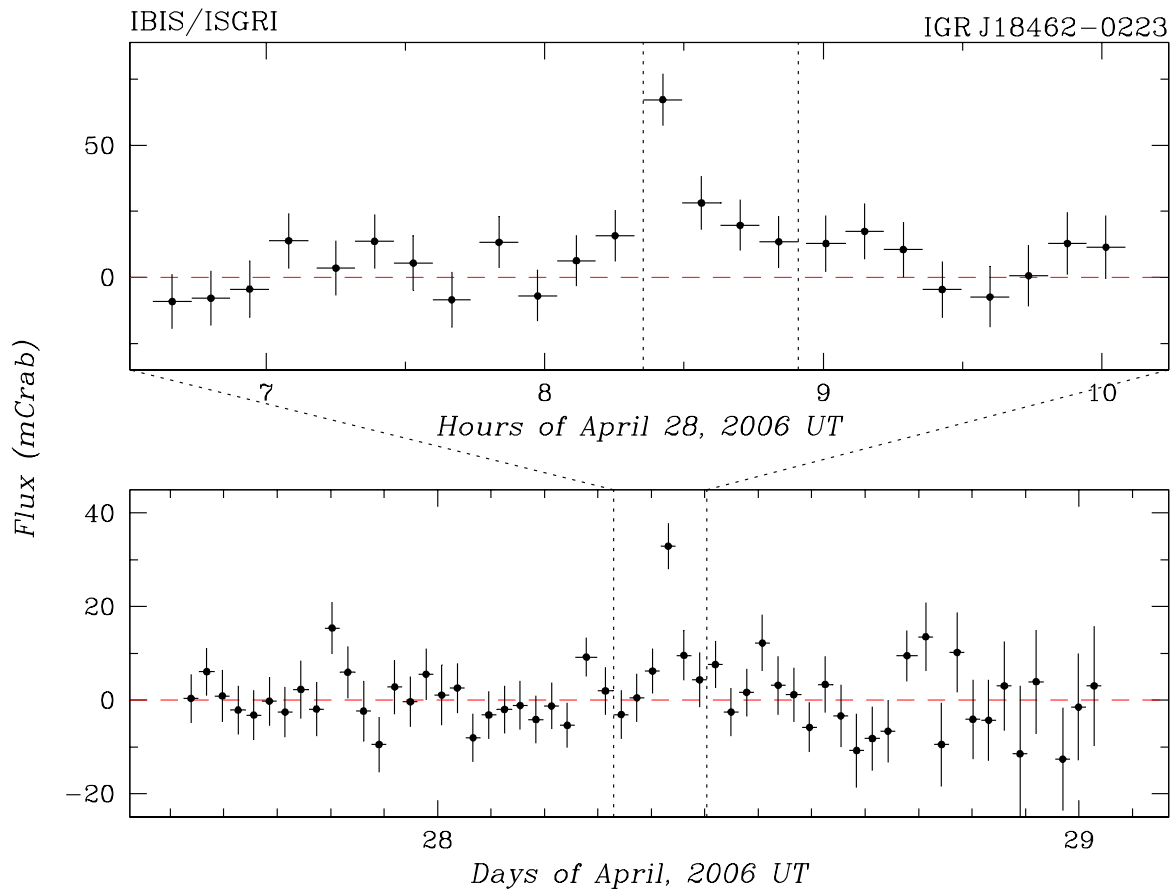}

\caption{\rm Light curve of \igr\ in the 20--60 keV energy band
  obtained by IBIS/ISGRI on April 27--28, 2006 (at the
  bottom). Each point of this curve corresponds to a separate
  INTEGRAL pointing $\sim2000$ s in duration. The segment of the
  curve near the source's outburst (in the interval marked by
  the vertical dotted lines) is shown at the top with a 500-s
  resolution.  The dotted lines in the upper panel indicate the
  interval used to measure the outburst spectrum.}
%
\end{figure}

\section*{RESULTS}
\noindent
The lower panels in Figs. 1 and 2 show the light curves of
\igr\ obtained by the IBIS/ISGRI telescope in the 20--60 keV
energy band on April 27--28, 2006, and October 6--17, 2007,
respectively.  In these time intervals, the outbursts of its
hard radiation were detected\footnote{Note that the ASM monitor of the
  RXTE observatory did not see the source at the time of the
  outbursts.}. The outburst profiles with an improved time
resolution are presented in the upper panels of the
figures. Whereas the first outburst has a total duration of
$\sim1$ h, being characterized by an intense ($\sim65$ mCrab)
narrow ($\la20$ min) initial emission peak, the second outburst
has a duration of $\sim5$ h at an almost flat profile
corresponding to a flux of $\sim35$ mCrab with evidence of a
beginning decline by the end of the observations. The detection
confidence ($S/N$ ratio) is $\simeq 7.3$ and $9.5$,
respectively, for the first and second outbursts. The time
intervals indicated by the dotted lines in the upper panels of
the figures were used for its determination.

\begin{figure}[t]
\epsfxsize=17cm
\epsffile{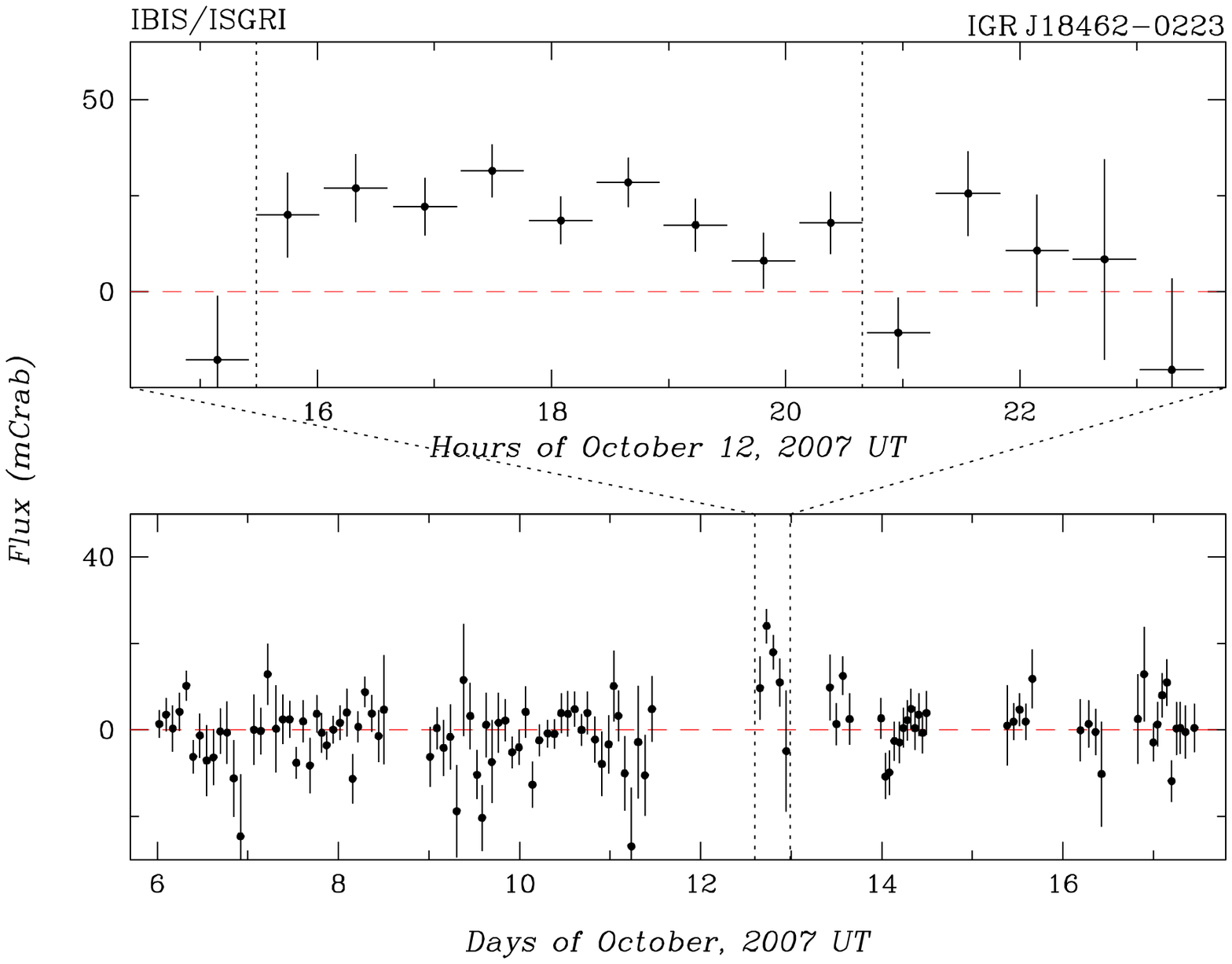}

\caption{\rm The same light curve as that in Fig. 1 but obtained
  in the period October 6--17, 2007, with a resolution of
  $\sim6000$~s (at the bottom). The profile of the outburst
  recorded in this period from \igr\ is shown at the top with a
  resolution of $\sim2000$~s corresponding to the duration of
  separate INTEGRAL pointings. The dotted lines in the upper
  panel indicate the interval used to measure the outburst
  spectrum.}
\end{figure}
Figure 3 presents the X-ray image ($S/N$ map) of the region near
\igr\ obtained by the IBIS/ISGRI telescope in the 20--60 keV
energy band from the sum of observations of the two outbursts.
The source was detected at a level of $S/N = 11.7$. Its position
on the map, R.\,A. $=18$\uh$46$\um$16$\,\fs$6$ and
Decl. $=-$$02$\deg$23$\arcmin$35$\arcsec\ (epoch 2000.0,
$1$\farcm$5$ uncertainty), slightly (by 24\arcsec) differs from
the position based on the data for only one (second) outburst
(Grebenev et al. 2007). We see from Fig. 3 that the source lies
in a densely populated region. This is natural, because it is
close to the Galactic plane and because the X-ray sources from
the Galactic arm (the base of the Scutum-Centaurus arm) are
projected onto this
region. It is quite likely that \igr\ itself
is located in this arm; the distance to it $d$ is then $\simeq6$
kpc. Most of these sources were detected during the \mbox{ASCA}
survey of the Galactic plane (Sugizaki et al. 2001), although
the fast X-ray transient IGR J18483-0311 discovered previously
by the INTEGRAL observatory (Chernyakova et al. 2003; Molkov et
al. 2004), the unique `schizophrenic' pulsar PSR J18464-0258 in
the supernova remnant Kes 75 (Kuiper and Hermsen 2009), and the
transient pulsar A 1845-024/GS 1843-02 (Finger et al. 1999) are
also seen here. None of the sources in the vicinity of
\igr\ indicated in Fig. 3 was detected in the image accumulation
time. The closest detected source, the well-known 361-s X-ray
pulsar XTE J1855-026, is at a distance of $\sim2$\deg\ from
\igr. It should be specially noted that AX J184600-0231 closest
to IGR J18462-0223 is 9\arcmin\ away from it, which exceeds
noticeably the uncertainty in localizing this new transient.
\begin{figure}[ht]
\epsfxsize=16.4cm
\epsffile{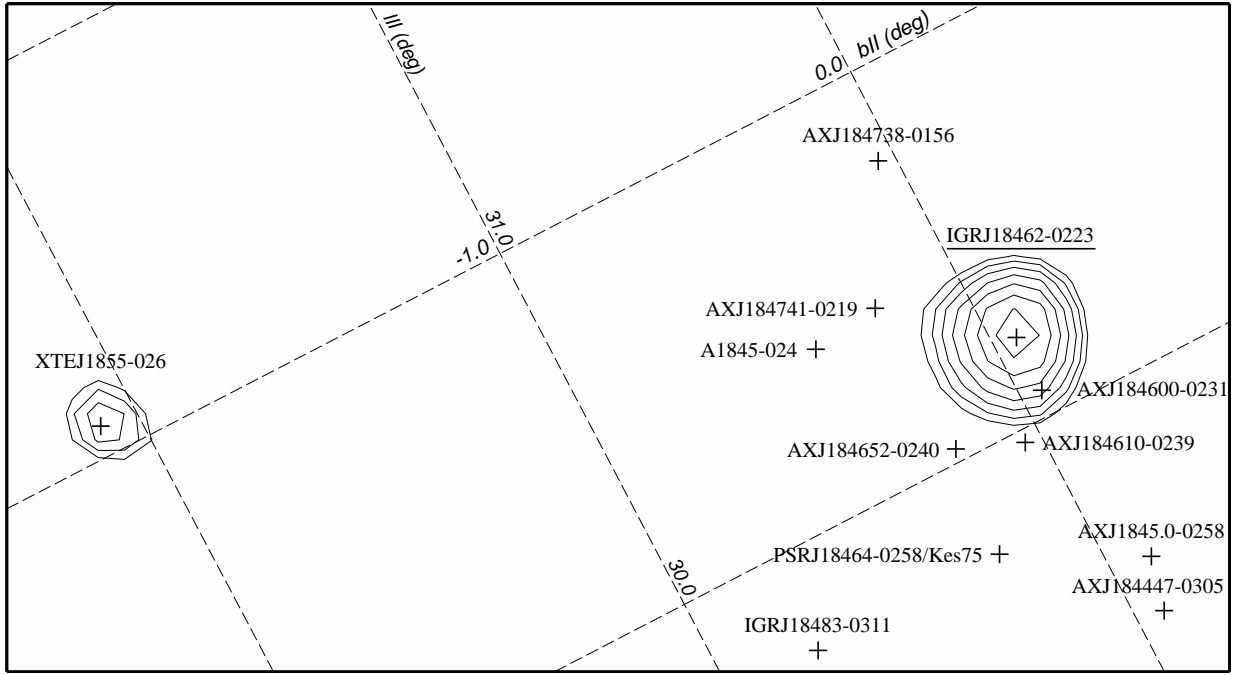}

\caption{\rm Image of the region near the X-ray transient
  \igr\ obtained by the IBIS/ISGRI telescope from the sum of
  observations of the 2006 and 2007 X-ray outbursts. The
  contours indicate the regions of confident detection of the
  sources in the energy range 20--60 keV and are given at $S/N$
  ratios of 3.5, 4.2, 5.0, 6.0, 7.2, 8.7, 10.4, 12.5, ... (on
  a logarithmic scale).  The image size is approximately
  $1$\fdg$5\times3$\deg.}
\end{figure}

Figure 4 shows the spectrum of \igr\ obtained from the sum of
observations of the 2006 and 2007 outbursts and the spectra of
each individual outburst. These were measured during the time
intervals indicated in Figs. 1 and 2 by the vertical dotted
lines. The corresponding average luminosities during the outbursts
calculated for an assumed distance of 6 kpc are listed in the
table. We see that the source's emission is detected at least up
to $\sim 80$ keV; a gradual drop is observed at energies above
$60$ keV. The spectra of the two outbursts are similar; the
first is harder than the second but mainly through the energy
range $\ga 60$ keV, where the points of the spectrum are
statistically not very significant.  Fitting the average
spectrum of the outbursts by the law of bremsstrahlung from an
optically thin thermal plasma (see the table, the TB model)
yields a temperature $kT\sim40$ keV. Spectra of such shape and
hardness are typical of X-ray pulsars in high-mass Xray binaries
and, in particular, fast X-ray transients.  To illustrate this
assertion, we provided the spectrum of the well-known fast X-ray
transient IGR J16479-4514 (taken from Grebenev 2009), which is
actually very similar to that of \igr\ with the same temperature
$kT\simeq38\pm6$ keV, in the upper panel of Fig. 4.
\begin{figure}[p]
\epsfxsize=10.5cm
\hspace{2.5cm}\epsffile{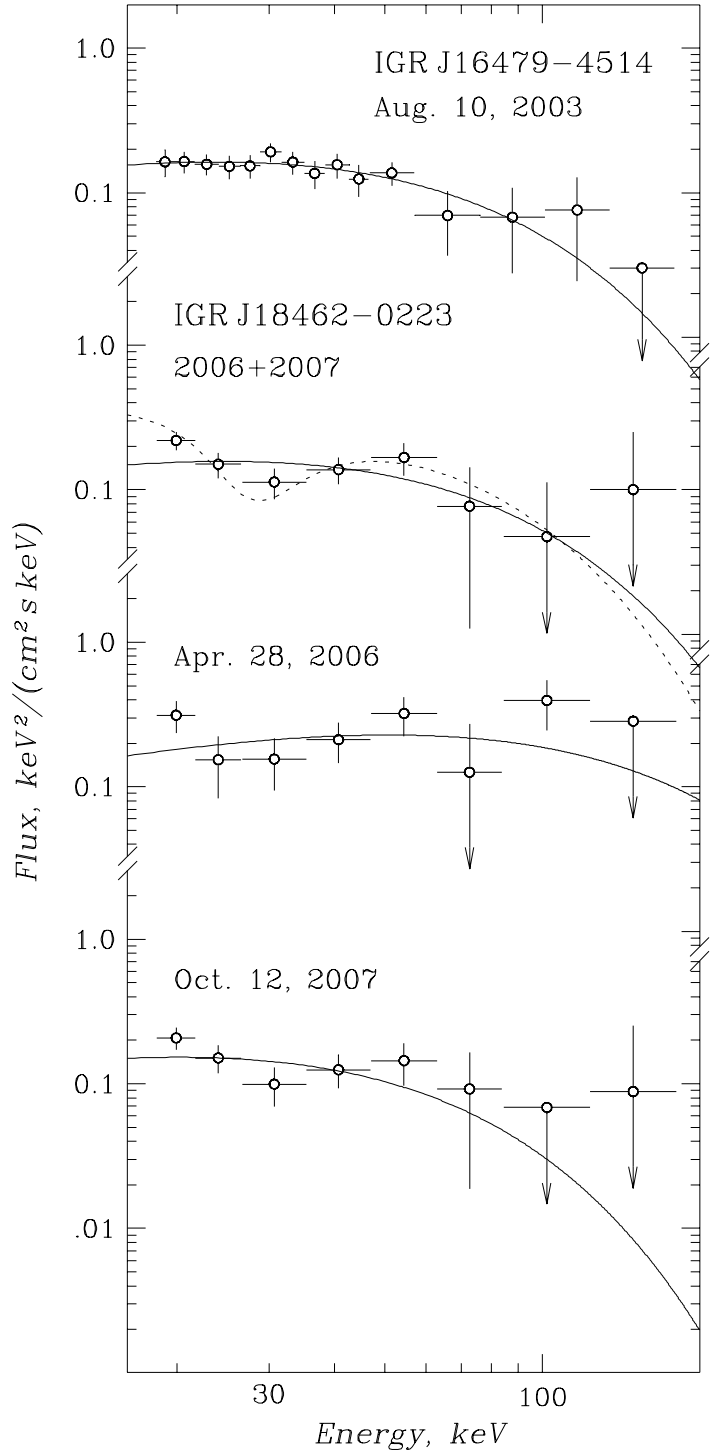}
\caption{\rm Average X-ray spectrum of \igr\ obtained by the
  IBIS/ISGRI telescope from the sum of observations of the 2006
  and 2007 outbursts and spectra of each individual
  outburst. The spectra are hard, with the characteristic
  temperature being $kT\simeq 30-80$ keV (solid lines) when
  fitted by the law of bremsstrahlung from an optically thin
  plasma. The dotted line indicates the result of including the
  cyclotron absorption line in the fit (see the text). For
  comparison, the spectrum of the well-known fast X-ray
  transient IGR J16479-4514 is shown at the top.}
\end{figure}
\begin{table}[t]

\vspace{6mm} 

\centering 

{{\bf Table.} Results of fitting the
  X-ray spectrum of \mbox{IGR\,J18462-0223}\protect\\ during the
  outbursts detected by the INTEGRAL observatory.}

\vspace{5mm}\begin{tabular}{c|c|c|c|c|c} \hline\hline
Date  &Model\a\ &$kT,$&$\alpha$\b\ &$L_{\rm X},$\c\ &$\chi^2_N (N)$\d\ \\ 
           &      & keV&       &$10^{36} \ \mbox{erg s}^{-1}$& \\ \hline
Apr. 28, 2006 &PL   &$            $&$2.08\pm0.37$&$2.65\pm0.38$ &1.18 (23)\\
           &TB   &$ 81\pm28    $&             &$2.75\pm0.40$ &1.22 (23)\\ 
&&&&&\\
Oct. 12, 2007 &PL   &             &$2.66\pm0.35$ &$1.40\pm0.15$ &1.39 (23)\\
           &TB   &$ 32\pm4$    &              &$1.37\pm0.15$ &1.47 (23)\\
&&&&&\\

2006+2007  &PL   &$            $&$2.48\pm0.28$&$1.61\pm0.15$ &1.28 (23)\\
           &TB   &$ 39\pm5     $&             &$1.60\pm0.14$ &1.39 (23)\\
           &TB+CA&$ 30\pm7     $&             &$1.61\pm0.14$ &1.07 (21)\\ \hline

\multicolumn{6}{l}{}\\ [-3mm]

\multicolumn{6}{l}{\a\ PL --- power law, TB --- thermal bremsstrahlung,}\\
\multicolumn{6}{l}{\ \ \ CA --- cyclotron absorption (one harmonic)}\\
\multicolumn{6}{l}{\b\ photon index}\\
\multicolumn{6}{l}{\c\ 20--100 keV luminosity  for an assumed
  distance of  $d=6$ кпк}\\ 
\multicolumn{6}{l}{\d\ $\chi^2$ value of the best fit normalized to
   $N$\ ($N$ is the number of degrees of
  freedom)}\\ 
\end{tabular}
\end{table}

Note that the average spectrum of the outbursts can be fitted by
a power law (PL in the table) with a photon index
$\alpha\simeq2.5\pm0.3$ as successfully as by a thermal
bremsstrahlung law. In general, however, a good fit (with $\chi^2$ 
normalized to the degree of freedom $\simeq1$) cannot
be achieved with such simple models, because there is a feature
in absorption near $\sim30$ keV or, possibly, an additional soft
emission component at energies $\la25$ keV in all spectra. This
feature can result from the resonance cyclotron absorption of
emission in a neutron star's strong magnetic field (see, e.g.,
Tsygankov et al. 2007). Adding a cyclotron harmonic in the form
of a factor,
$$\exp\left[-\tau_c\left(\frac{E}{E_c}\right)^2 \frac{\Delta
    E_c^2}{(E-E_c)^2+\Delta E_c^2}\right],$$ to the
bremsstrahlung model actually improves noticeably the fit (see
the table), although this line cannot be said to be present with
certainty due to the overall low statistical significance of the
spectrum points (according to the $\Delta\chi^2$-statistics the
probability of the line appearance by chance is
$8\times10^{-3}$). The width of the cyclotron line was fixed at
its best value of $\Delta E_c=8$ keV, with its optical depth and
energy being $\tau_c=1.4\pm0.4$ and $E_c\simeq 26\pm2$ keV,
respectively. The value of $E_c$ obtained allows the neutron
star's magnetic field in the region of main energy release to be
estimated, $B= (1+z)\, E_c/( 11.6\ \mbox{\rm
  keV})\ \times10^{12}\ \mbox{\rm G}\simeq
2.2\times10^{12}\ (1+z)$ G, where $z$ is the redshift in the
neutron star's gravitational field.

\section*{CONCLUSIONS}
\noindent
The short (several hours), intense X-ray outbursts of \igr,
their hard spectrum similar to the spectrum of bremsstrahlung
from an optically thin plasma with $kT\sim 30$--$40$ keV, the
long (years) intervals between the outbursts, and, finally, the
source's location in the Galactic plane and, possibly, in the
Galactic arm, where intensive star formation takes place, all
allow us to consider \igr\ as a new candidate for fast X-ray
transients with a massive companion or even a supergiant. Its
more accurate localization (in soft X-rays) and optical
identification are needed. Hard X-ray observations of new
outbursts from \igr\ will allow the question about the presence
of a cyclotron absorption line in its spectrum to be solved. The
number of pulsars whose radiation spectra exhibit such lines is
still small, while the importance of such observations, which
allow direct measurements of a neutron star's magnetic field to
be made, is great.

Note that the outbursts of \igr\ are shorter than those for most
other fast transients.  What is this --- a characteristic
feature of this source or an observational effect related to its
fairly large distance ($d\sim6$ kpc) and its location at the
edge of the telescope's field of view during the outbursts, in
the region where the effective area of the telescope decreases?
The measured flux from \igr\ during the outbursts was comparable
to that from such fast transients as IGR J16465-4507, IGR
J16479-4514, and several others, while its detection confidence
was lower. Since the profiles of outbursts from many fast
transients contain several individual intense bursts standing
out above the average level (see the selection of light curves
for fast transients in Grebenev 2009), the possibility of
observational (selective) effects due to which we see only the
tops of such bursts in the case of \igr\ seems quite
real. Therefore, the following question is natural: How many
unknown fast transients similar to \igr\ still wait for their
`lucky' (direct and timely) hard X-ray telescope pointing to be
detected?

\section*{ACKNOWLEDGMENTS}
\noindent  This work is based on the observational data
from the INTEGRAL observatory\footnote{An ESA satellite with
  scientific instruments provided by France, Italy, Germany,
  Switzerland, Denmark, Spain, Czech Republic, Poland,
  launched in orbit by Russia, and operated by ESA with
  participation of USA.} retrieved via the Russian and
European INTEGRAL Science Data centers. We used some of the
programs developed by E.M. Churazov to analyze the data. We are
grateful to R.A. Krivonos for access to the results of the
all-sky survey carried out by him on the basis of INTEGRAL data. The
study was supported by the Russian Foundation for Basic Research
(project no. 10-02-01466), the Program of the Russian President
for Support of Leading Scientific Schools (grant no.
NSh-5069.2010.2), and the `Origin and Evolution of Stars and
Galaxies' Program of the Presidium of the Russian Academy of
Sciences.\\
 
\pagebreak   


\hfill {\it Translated by Shtaerman}
\end{document}